\documentstyle[12pt,epsf,epsfig,wrapfig]{article} 
\newcommand{\conj}[1]{\buildrel \ast \over{#1}}
\begin{document}
%
\baselineskip=\normalbaselineskip
\begin{center}
{\Large{\bf 
\strut POLARIZATION PROPERTIES  \\
\strut OF DIFFRACTIVELY PRODUCED $\Lambda_c^+$}}\\[2mm]
{\bf Yu. Arestov$^1$ and F.R.A. Sima\~{o}$^2$}\\[1mm]
\end{center}
%
\parbox{\textwidth}{
{\it 1 - Institute for High Energy Physics, 142284 Protvino, 
Russia}\\
{\it 2 - Centro Brasileiro de Pesquisas Fisicas (CBPF), R. Xavier Sigaud 150,\\
 22290-180, Rio de Janeiro, Brazil}\\[3mm]
}
\begin{center}
\parbox{0.85\textwidth}{
    The Pomeron-gluon-gluon interaction is considered in the
    QCD-based model for the charmed baryon production in the process
    ${\rm Pomeron} \,+\, p \to \Lambda_{c}^+ \,+\, X$. 
    The polarization of the produced heavy quark is induced effectively 
    through the non-perturbative long-range interaction with the gluon
    field of the type $[\vec \sigma \cdot {\rm rot} \cal A ]$.
    The $x_F$-dependence of $\Lambda_{c}^+ $ polarization, 
    $P_{\Lambda_{c}^+}(x_F,p_T)$,
    has been studied. Its absolute value  depends on the model parameter
    $a$ and it appears to be sizeable in the wide range of $a$
    values: when $a$ ranges from 0.1 to 1.0, the polarization 
    $ P_{\Lambda_{c}^+}(x_F,p_T)$ varies from --0.2 to --0.5 at 
    $x_F \sim 0.5$
    and $p_T$ in the interval 1$\div$2\,GeV/c.
}
\end{center}
%
%
{\bf Introduction.} Polarization properties of heavy quarks have been studied
for a long period, both experimentally and theoretically. A major piece of
knowledge came from the experimental study of strange $\Lambda$ baryons.
The polarization of the final-state $\Lambda$'s at large $x_F$ is negative 
ant it reaches 20\,-\,30\%. The most recent and full review on the 
polarization of the strange baryons can be found in K.\,Heller's report
at SPIN96 \cite{Heller}.

   So far, the experimental data on the charmed $\Lambda_c^+$ polarization 
have been restricted to the output of only one experiment 
performed in IHEP, Protvino at the neutron beam
in the reaction $n\,+\, {\rm C_{12}} \to \Lambda_c^+ \,+\, X$ \cite{Lubimov}. 
The measured upper limit of the absolute value of 
$\Lambda_c^+$ polarization was equal to 0.5\,$\pm$\,0.2\,. The momentum of
the neutron beam ranged from 40 to 70 GeV/c.

   Perhaps soon new data on the $\Lambda_c^+$ polarization will come
from the E-831 experiment at Fermilab (FOCUS) in which the polarization
measurements can be performed in the reaction 
$\gamma \, + \, p \to \Lambda_c^+ \, + \, X$ at 300 GeV/c.

   The forthcoming collider experiments at RHIC and LHC will make it
possible to study the $\Lambda_c^+$ production and its spin properties in
various kinematic regions, in particular, in the diffraction region.
Theoretically the spin phenomena in diffractive processes are under intensive
study. Three refs. in \cite{close} represent different aspects of the spin
dependence of diffractive interactions.

   Unlike the strange $\Lambda$, the theory predictions for the 
$\Lambda_c^+$ polarization $P$ are not numerous. The large values up to 100\%
were predicted for the absolute value of $P_{\Lambda_c^+}$ in an early paper 
\cite{YuA1}.
The recent estimates were made in ref. \cite{Simao} for the intrinsic
charmed sea contribution to the $\Lambda_c^+$ production. Under  various
model assumptions, $P_{\Lambda_c^+}$ ranges from  --0.02 to --0.11  at large 
$x_F$ in the recombination approach.

   Note also that the spin of $\Lambda_c^+(2285)$ has not firmly been 
established 
so far. In the following we shall assume that the spin is equal to 1/2
as it is accepted by the PDG \cite{PDG}.

   In this paper we are interested in the $\Lambda_c^+$ polarization
in the diffractive production process
\begin{equation}
         p + p \to \Lambda_c^+ + X   \label{eq1}
\end{equation}
at high energies (fig.\,1). In the upper block the $c \bar c$ pair is
produced and $c$ quark is polarized in the final state. The model for
the heavy quark polarization depends on the kinematics in the upper block 
and  does not promptly depend on the diffractive production of the upper
block. The advantage of the diffractive $c \bar c$ production in (\ref {eq1}) 
displayed in fig.\,1 is the simplicity 
of the lowest-order model with the dominance of the Pomeron-gluon-gluon 
coupling. The more general consideration would require an additional 
determination of, at least,
the ratio of the coupling constants Pomeron-gluon-gluon/Pomeron-quark-quark.

  {\bf Formulation of the model.}  We study the inclusive production of 
$c \bar c$ pair by the initial Pomeron in the process
\begin{equation}
         p + {\rm Pomeron} \to \Lambda_c^+ + X   \label{eq2}
\end{equation}
(fig.\,2) which is relevant to the inclusive reaction (\ref{eq1}).
In this reaction the incoming proton energy  $\sqrt{s}/2$ and the Pomeron 
4-momentum  $q$ are those of reaction (\ref {eq1}). $q$  is fixed
at an arbitrary small value. The results of the calculations of the
polarization $P_{\Lambda_{c}^+}(x_F,p_T)$ do not depend 
essentially on $q$ in the diffractive region.

   The underlying parton subprocess is shown in fig.\,2 as the lowest-order
heavy quark pair production by the gluon-Pomeron coupling:
\begin{equation}
         g + {\rm Pomeron} \to c\,+\, \bar c   \label{eq3}.
\end{equation}

  We neglect the charmed sea, so the leading order
process $c \,+\, {\rm Pomeron} \to c \,+\, g$ does not contribute to the 
$\Lambda_c^+$ polarization. Also we neglect parton transverse momenta.
The $\Lambda_c^+$ transverse momenta in (\ref{eq2}) are taken in the region 
$p_T \,=\, 1 \div 2$ GeV/c.

   Another assumption is the dominance of the Pomeron-gluon-gluon coupling
constant  $g_{Pgg}$ compared to the Pomeron-quark-quark coupling $g_{Pqq}$,
so that $g_{Pqq}$/$g_{Pgg} \ll \, 1$. Hence, at this level we exclude from
the consideration the next-order subprocess 
$q \,+\,{\rm Pomeron} \to q \,+\, c \bar c$ in which a valence quark $q$ 
is excited by the Pomeron and it emits the $c \bar c$ pair.

   The produced $c$ quark recombines inclusively with the spectator $ud$ 
diquark system thus forming the final $\Lambda_c^+$.

   The $\Lambda_c^+$ polarization originates from the polarization of the
produced $c$ quark and is expressed in terms of the quark spin density
matrix $ \rho_Q^{\lambda \lambda '}$ by the following formula for the 
charmed baryon spin density matrix in  process (\ref{eq2}):
\begin{eqnarray}
\rho^{\lambda \lambda '}\,E\frac {d^3 \sigma}{dp^3} =%
\int {\rm d}x_{ud} {\rm d}x_g {\rm d}z\, f_g(x_g,Q^2)&\,f_{ud}(x_{ud},Q^2)\,
\varepsilon_Q \frac{d^3 \hat \sigma}{dp_Q^3}\,
 \rho_Q^{\lambda \lambda '}(\vec p_Q, {\cal A})\, \nonumber\\
 &  \times R(x_F,p_T,x_Q,x_{ud};z).  \label{rhol}
\end{eqnarray}
Here $f_g$ and $f_{ud}$ are gluon and diquark distribution functions with
$x_g$ and $x_{ud}$ standing for the momentum fractions of the  gluon and
diquark, respectively. $Q^2$ is the QCD scale chosen in the form 
$Q^2$\,=\,$m_Q^2\,+\,p_T^2$. The quantities $\hat \sigma,\,p_Q,\,
\varepsilon_Q$ refer to the underlying subprocess (\ref{eq3}). The 
hadronization process in (\ref{rhol})
is described by the inclusive recombination function $R$ depending on the 
$c$ quark, diquark and $\Lambda_c^+$ momenta with $z$ being the energy
fraction of $(ud,c)$ system carried away by $\Lambda_c^+$.
   Using (\ref{rhol}) the $\Lambda_c^+$ polarization is written as 
$ P_{\Lambda_c^+}\,=\,{\rm Tr}(\sigma \rho)/{\rm Tr}(\rho)$.

Initially a model for the final-state quark polarization expressed in terms of 
$ \rho_Q^{\lambda \lambda '}(\vec p_Q, {\cal A})$  was 
formulated in \cite{YuA2}. In this paper we explore a corrected version of 
that model applied to the charmed baryon production.
In the model the produced heavy quark scatters off
the external gluon 
field ${\cal A}$ and the quark spin $\vec s_Q$ may couple to the field, so that
the correlation $\vec s_Q \cdot {\rm rot} {\cal A}$ occurs. 

   To estimate this correlation, a non-trivial model for the external gluon
field ${\cal A}$ itself is needed. In case of the simplest  color
field, the spin flip properties and polarization of a scatttered massive quark 
were considered, for example, in the early paper \cite{Szwed} by J.\,Szwed,
where the external gluon field $\Phi^a(q)=4\pi g I^a/\vec q^2$ was used
in the perturbative second-order calculations.
Note that the massive quark polarization obtained in \cite{Szwed} was 
negative.

An alternative way is  to parametrize the heavy quark spin density matrix
 $ \rho_Q ^{+-}$ globally rather than to introduce an explicit model for
${\cal A}$ and for the quark interaction with this field.

   The parametrization of $\rho_Q$ is based on the following considerations. 
By definition $\rho_Q^{\lambda \lambda '}\,=\, 
\sum L^{\lambda \mu} \conj{L}{^{\lambda ' \mu}}$ where $L^{\lambda \mu}$ is 
the quark transition amplitude in the gluon field ${\cal A}$.
The following traces relate to the basic formula (\ref{rhol}):
${\rm Tr}(\sigma \rho_Q)\,=\,-2{\rm Im}\,\rho_Q^{+-}\,=\, 
4{\rm Im}\,(L^{++}\conj{L}{^{+-}})  $ 
and ${\rm Tr}(\rho_Q)\,=\,2(\vert L^{++}\vert^2 \,+\, \vert L^{+-}\vert^2)$.
Under the space parity conservation the relations $L^{++}=L^{--}$ and 
$L^{+-}=-L^{-+}$ hold.

    For the parametrization it is convenient to include ${\rm Tr}(\rho_Q)$
into the recombination (hadronization) function $R$ in (\ref{rhol}). Then
the trace of the convolution of (\ref{rhol}) with the Pauli matrix 
$\vec \sigma$ will contain in the integrand the expression 
$P_Q\,=\,{\rm Tr}(\sigma \rho_Q)/{\rm Tr}(\rho_Q)$ 
which is exactly the heavy quark polarization (see below).

   After these generalities the time has come to parametrize the ratio $P_Q$.
The normalized spin-flip transition amplitude squared 
$\vert L^{+-} \vert ^2$ 
has the sense of probability to make 
quark spin-flip in the external field. 
It seems natural to think that it 
increases with the quark mean range.
Henceforth, it grows with increasing 
quark energy, $\varepsilon _Q$, since at hadronization the mean range of a 
fast quark is proportional to $\varepsilon_Q /M^2$ where $M^{-1}$ is of order
of the confinement radius. Thus, we accept the equality 
$\vert L^{+-} \vert ^2 /(\vert L^{++} \vert ^2 + \vert L^{+-} \vert ^2)\,=\,
a \varepsilon_Q /\varepsilon_Q^{max}$ with the supression parameter
0$<a<$1. The normalization $\varepsilon_Q^{max}$ depends, of course, on the
kinematics, however, for the simplicity we take it equal to $\sqrt{s}$/2.

  Taking these straightforward considerations into account the parametrization
 has been made as follows:
\begin{equation}
\nonumber
         P_Q \,=\,
         \frac{2{\rm Im}(L^{++}\conj{L}{^{+-}}) }
          {\vert L^{++} \vert ^2 + \vert L^{+-} \vert ^2} =
         2\sqrt{ax_{\varepsilon}(1-ax_{\varepsilon})}
          \sin \Delta \phi,  \nonumber         \label{qpol}
\end{equation}
where $x_{\varepsilon}=\varepsilon_Q /\varepsilon_Q^{max}$, and
$\Delta \phi$ is the phase difference of the amplitude $L^{++}$
and $L^{+-}$.

 {\bf Phase difference.} In principle the phase difference $\Delta \phi$
depends on arguments of the amplitudes $L^{\lambda \lambda '}$, i.e. on
$\varepsilon_Q$. The simple linear dependence 
$\Delta \phi = r_1 \varepsilon_Q + r_2$ was used in \cite{YuA3} to describe
the experimental data on the one-spin asymmetry in the reactions 
$p_{\uparrow} + p \to \pi^{\pm} + X$ at 13 and 18 GeV/c and in the reaction
$\pi^- + p_{\uparrow} \to \pi^{\circ} + X$ at 40 GeV/c. Three sets of 
experimental data were enough to determine the parameters $r_1$ and $r_2$.

   In this paper we prefer to imply $\Delta \phi$ value based on the 
Regge-type considerations. The Regge properties of the reaction (\ref{eq2}) 
are determined by $D^{\ast}$ and $D^{\ast \ast}$ exchange trajectories 
(see insertion in fig.\,1). And in the Regge approach the $\Lambda_c^+$
polarization in reaction (\ref{eq2}) is due to the interference of the 
exchange diagrams with these trajectories \cite{YuA1}. 
Thus the difference of the Regge phases,
$\Delta \phi = \phi_{D^{\ast}} - \phi_{D^{\ast \ast}}$ occurs naturally.
It depends on $t$, the squared momentum transfer from the initial proton
to the final charmed baryon. Since we consider $\Lambda_c^+$ production
at large $x_F$ and relatively small $p_T$, we neglect the small $t$
contribution in $\Delta \phi$ (in \cite{YuA1} the $t$-dependence dropped
down due to the hypothesis of the weak exchange degeneration). So, we obtain 
$\Delta \phi = \pm (j_{\ast}-j_{\ast \ast})\pi /2 = \pm \pi /2$. Here
the starred $j$'s correspond to the spins of the exchange trajectories.
The sign $\pm$ reflects the non-uniqueness of this procedure and it can be
chosen by comparing the results with the strange $\Lambda$ polarization
at large $x_F$.

{\bf Parton distributions.} We used two sets of the leading-order parton
distribution functions (pdf's): CTEQ4L and GRVLO \cite{pdf}.
In the kinematic region of interest, the gluon pdf's $f_g(x,Q^2)$ from these
sets are close at $x<0.3 \div 0.4$ and they differ at larger $x$, whereas 
the pdf's of the light quarks $u$ and $d$ are similar.

   For our purpose,  the 'diquark' pdf has been constructed of $u$ and $d$
quarks very simply:
$$ f_{ud}(x,Q^2)\,=\, 
   \int^x \frac{f_u(y,Q^2)}{y}\frac{f_d(x-y,Q^2)}{x-y}\, {\rm d}y\; . $$
The arguments in the denominators means that the pdf's output from the
packages in \cite{pdf} are of the conventional type $x$\,times\,pdf.
The diquark distribution is shown in fig.\,3.

   {\bf Inclusive recombination.} The hadronization of the produced charmed 
quark is assumed via the inclusive recombination with  the spectator valence 
$ud$ system. 
This subprocess is described by the inclusive recombination function 
$$R(x_F, p_T,x_Q,x_{ud};z)\,=\, const \cdot \frac{x_{\varepsilon_Q}x_{ud}}
{x_{\varepsilon_Q} + x_{ud}} \cdot \frac{1}{z} \delta (z-1) $$
where $x_E$ is the reduced $\Lambda_c^+$ energy: $x_E=2E/\sqrt{s}$, and
$z=x_E /(x_{\varepsilon_Q} + x_{ud})$.
This expression incorporating the transverse momentum is a generalization of 
the conventionally used recombination functions.
The normalization parameter $const$ has not been specified since it drops
down in the formula for polarization $P_{\Lambda_c^+}$.

   {\bf Partonic subprocess.} The differential cross section 
$\varepsilon_Q {\rm d}^3 \hat \sigma /{\rm d}p_Q^3$ has been calculated 
using the matrix element 
$${\rm M}\,=\, g_{Pgg}g_s \frac{1}{\hat s} \phi_P T_{ij}^a \epsilon_{\mu}^a 
 {\bar u}_Q^i \gamma_{\mu}v_{\bar Q}^j . $$
Here $g_{Pgg}$ and $g_s$ are the Pomeron-gluon-gluon and the strong coupling
constants, respectively, $\epsilon_{\mu}^a$ is the gluon polarization 
4-vector, $u_Q$ and $v_{\bar Q}$ are the quark and antiquark spinors,
respectively, $T_{ij}^a$ is the color factor and $\hat s$ is the incoming
energy squared. $\phi_P$ is the Pomeron wave function which is normalized
as $\phi_P \phi_P^{\ast}$=1.

   {\bf Kinematics.} We are interested in $x_F$ dependence of the 
$\Lambda_c^+$ polarization $P_{\Lambda_c^+}(x_F,p_T)$ 
in the process (\ref{eq2}) with given $p_T$ and 
at a fixed value of $q$. The 
kinematic region 0.2$ <x_F <$0.9 and 1$<p_T<$2\,GeV/c was studied. No visible
$p_T$ dependence of $P_{\Lambda_c^+}(x_F,p_T)$  was obtained.

  The basic relations for the longitudinal and transverse momenta are the 
following: $x_F=x_{ud}+x_Q$, 
$\vec p_T=\vec p_{TQ}$. This should be added by the 4-momentum conservation
in the partonic subprocess. Thus, the integration limits in $x_{ud}$ and 
$x_g$ can be formally determined. However the lower limit in $x_{ud}$, the
momentum fraction of the spectator valence diquark would be too small 
from the physical point of view. So the lower limit of $x_{ud}$ was chosen
to be not less than 0.15. Variations around this value do not cause 
the strong change of the resulting polarization.

    The lower limit in integrating over $x_g$, the gluon momentum fraction,
is defined by $x_g^{min}=(2/\sqrt{s})(\sqrt{p_T^2 + m_Q^2}+\varepsilon_Q
-q_0)$. The value of the upper limit has been chosen as $x_g^{max}$=\,0.95.

  {\bf An important note.} Here is an appropriate place to make a note on
the seemingly numerous assumptions before starting to calculate the 
polarization integrals.
The final-state baryon polarization at large $x_F$ is due
entirely to the QCD twist-3 contributions. For the twist-3 studies in hadron
interactions, there is no commonly used model. At first glance, there is too
much freedom in the above considerations of the quark polarization $P_Q$ 
and/or the inclusive recombination function $R$. It seems to be a 
disadvantage of the
approach. However the hadron ($\Lambda_c^+$ in this paper) polarization is 
{\it a ratio} of two
integrals which include the same functions in the nominator and denominator,
except $P_Q$ presented in the nominator only. And all the functions are
rather smooth. From this it follows that the ratio of the integrals depends
weakly on the reasonable variations of the separate functions in the 
integrands. It means, for example,  that our results must not change strongly 
if we use the second order diagrams in 
$\varepsilon_Q d\hat \sigma /dp^3 $ instead of the 
leading order calculations. The same is true for different forms of the
inclusive recombination function $R$. The shape and size of the 
$\Lambda_c^+$ 
polarization are mostly determined by the parametrization of $P_Q$ in 
(\ref{qpol}).

   {\bf Results of calculations.} As is seen from (\ref{rhol}),
the sign of the $\Lambda_c^+$ polarization
is determined by the sign of Tr($\sigma \rho_Q$), that is the sign of the
heavy quark polarization $P_Q$. The presented model for $P_Q$ (\ref{qpol})
has the sign ambiguity mentioned above in the paragraph {\bf Phase 
difference}. Since the model makes no difference
between the flavors of the massive quarks, we resolve the sign ambiguity
in favor of the strange $s$ quark in order to obtain the negative $\Lambda$
polarization at large $x_F$ in similar calculations. So the phase difference
$\Delta \phi$  in (\ref{qpol}) has been chosen equal to --$\pi$/2. It is worth
to note that this choice supported by the experimental data for the strange
$\Lambda$ polarization coincides with the sign of $P_Q$ in \cite{Szwed}
obtained in perturbative calculations.

   The only undetermined quantity is still the suppression factor $a$ in 
formula (\ref{qpol}). The numerical calculations were made for two 
far-distant values $a$=0.1 (small polarization) and  $a$=1 
(largest polarization) using two sets of the leading-order
parton distributions CTEQ4L and GRVLO \cite{pdf}. The results for 
$x_F$ dependence of the $\Lambda_c^+$ polarization 
$P_{\Lambda_c^+}(x_F,p_T)$ at  $p_T$=\,1.5 GeV/c  are shown in fig.\,4.
At $x_F$=\,0.5 the small and the largest polarizations differ in 
magnitude by a factor of $\sim$2.5. With growing $x_F$ from 0.2 to 0.9,
both the small and the largest polarizations increase by a factor of 2.

   {\bf Acknowledgement.} We would like to thank Hung-Liang Lai, one of the 
authors of CTEQ4 pdf package \cite{pdf}, for the instructive help in 
using CTEQ.

\newpage
\vspace*{-1.5cm}
\begin{center}
\hspace*{1.5cm} \epsfig{figure=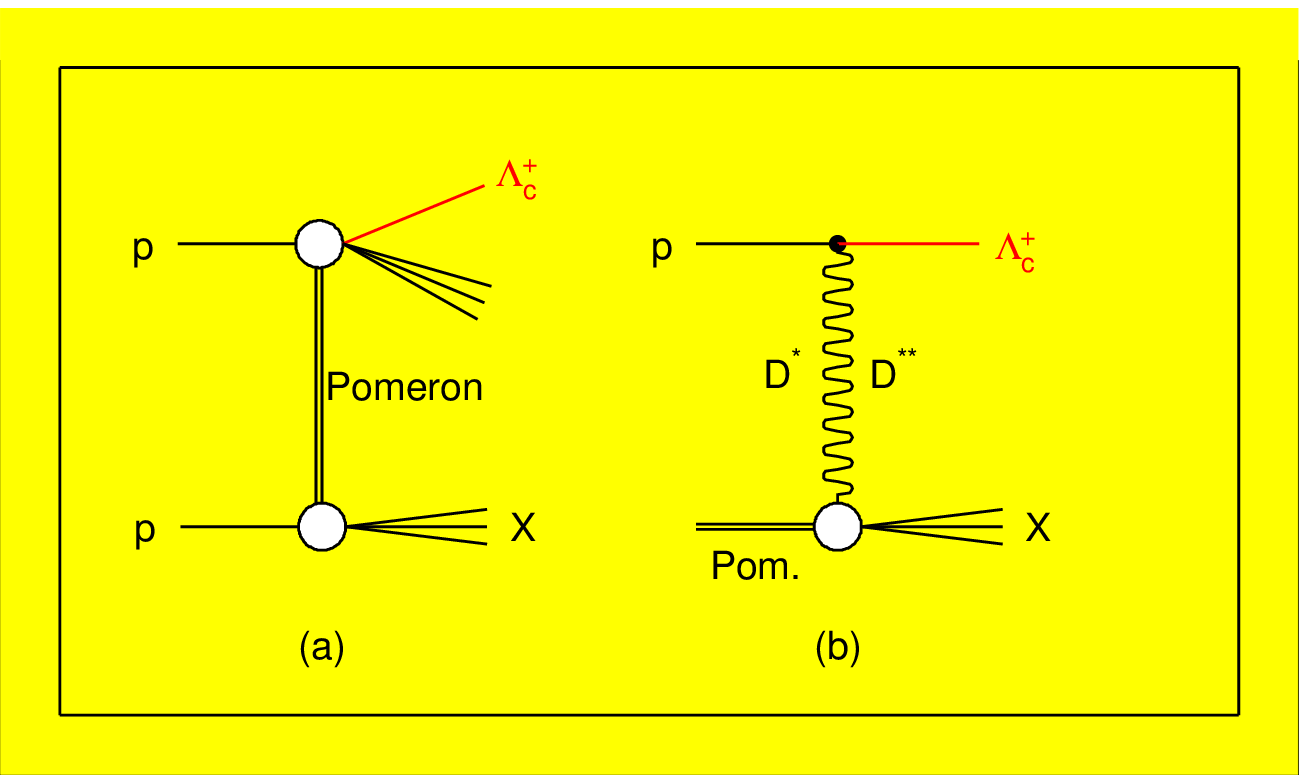,width=12cm,height=8cm}\\
\vspace*{4mm}
\hspace*{1.5cm} \parbox{12cm}{
  Fig. 1. (a) -- Diffractive production of $\Lambda_c^+$ with $c\bar c$ pair
in  the upper block in reaction (1); (b) -- Regge exchanges 
in the upper block.} \\[1cm]
\end{center}
%
%
\begin{center}
\hspace*{1.5cm} \epsfig{figure=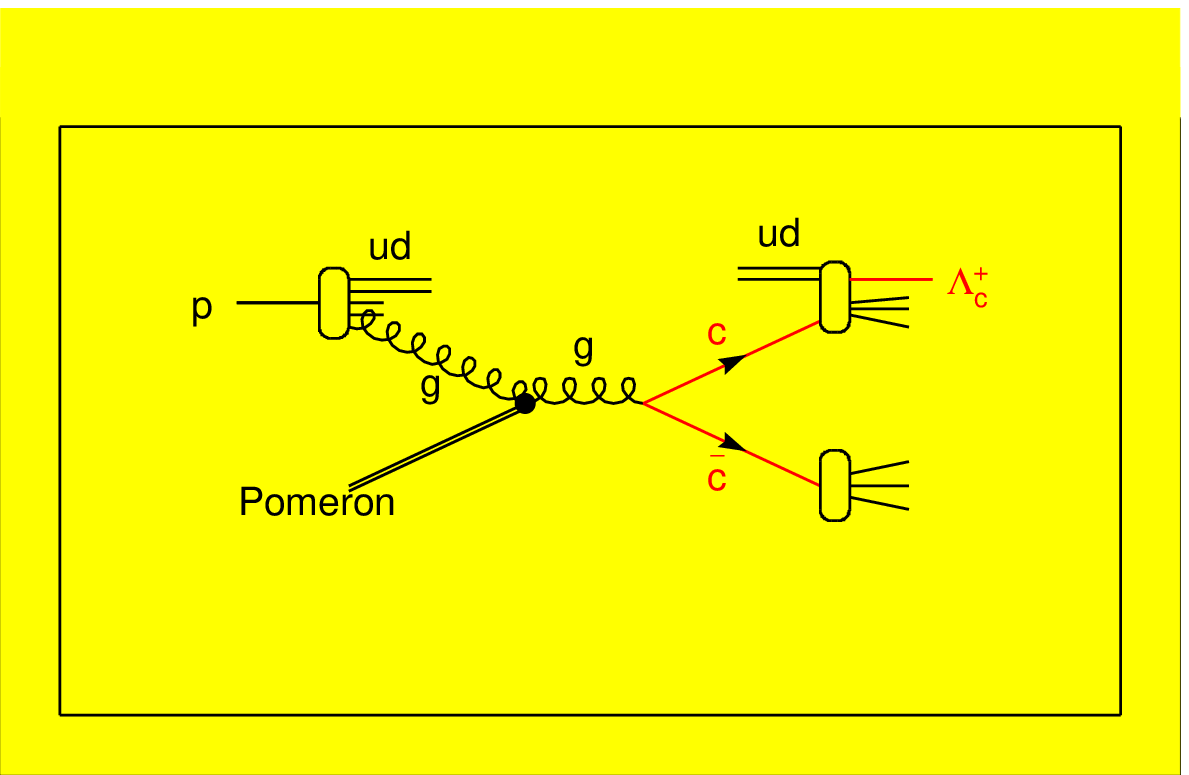,width=12cm,height=8cm}\\
\vspace*{4mm}
\hspace*{1.5cm} \parbox{12cm}{
  Fig. 2. The leading-order parton subrocess in reaction 
  $p\,+\,{\rm Pomeron} \to \Lambda_c^+ \,+\, X$. } 
\end{center}
\newpage
\vspace*{-1.8cm}
\begin{center}
\hspace*{1.5cm} \epsfig{figure=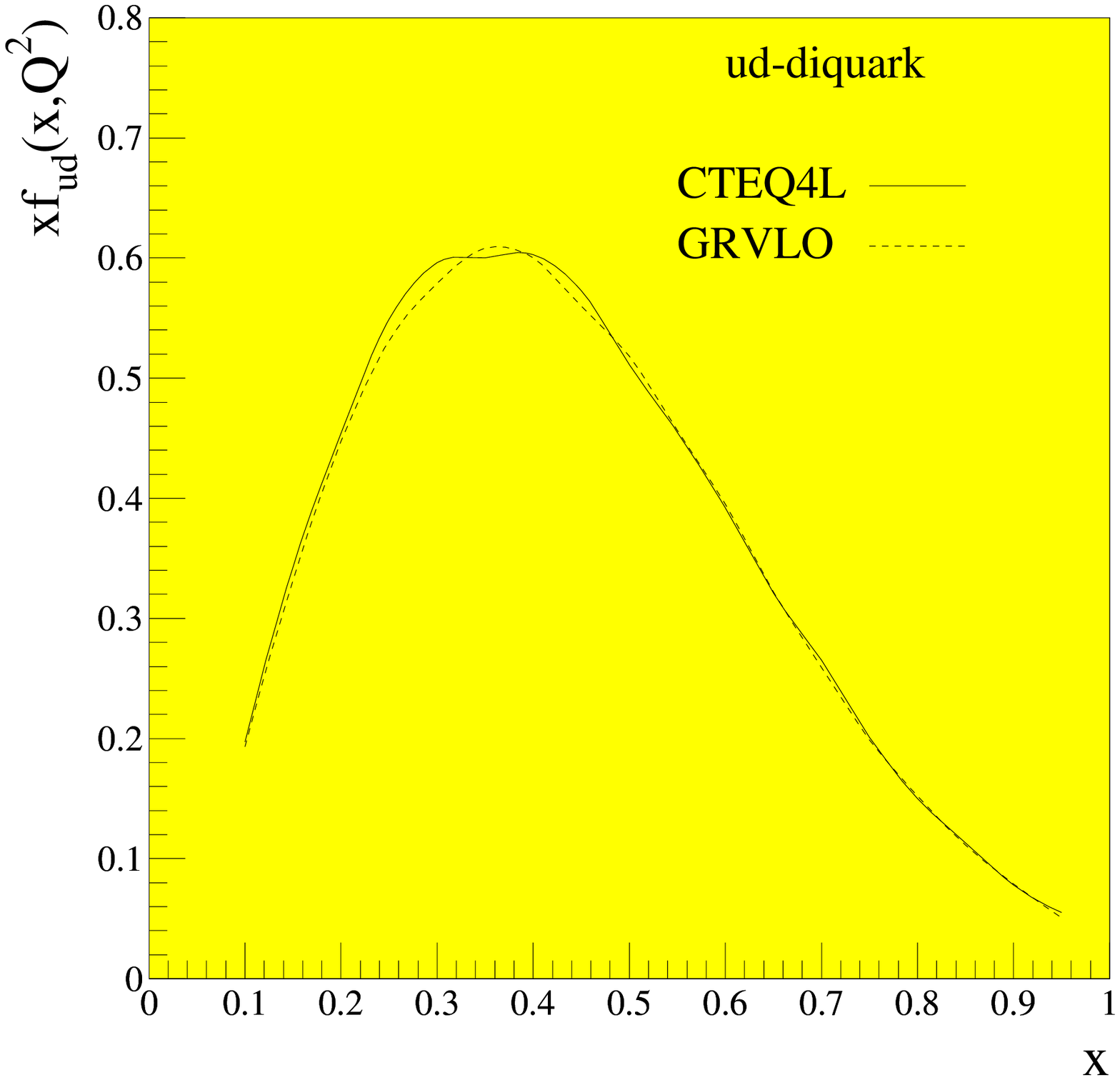,width=12cm,height=8cm}\\
\vspace*{4mm}
\hspace*{1.5cm} \parbox{12cm}{
  Fig. 3. Diquark distribution function $xf_{ud}(x,Q^2)$ at 
$Q^2\,=\,m_Q^2 \,+\, p_T^2 $ with $m_Q$\,=\,1.5\,GeV/c and 
$p_T$\,=\,2\,GeV/c. The solid (dashed) line corresponds to CTEQ4L(GRVLO)
parton distribution functions.} 
\end{center}
%
%
\begin{center}
\hspace*{1.5cm} \epsfig{figure=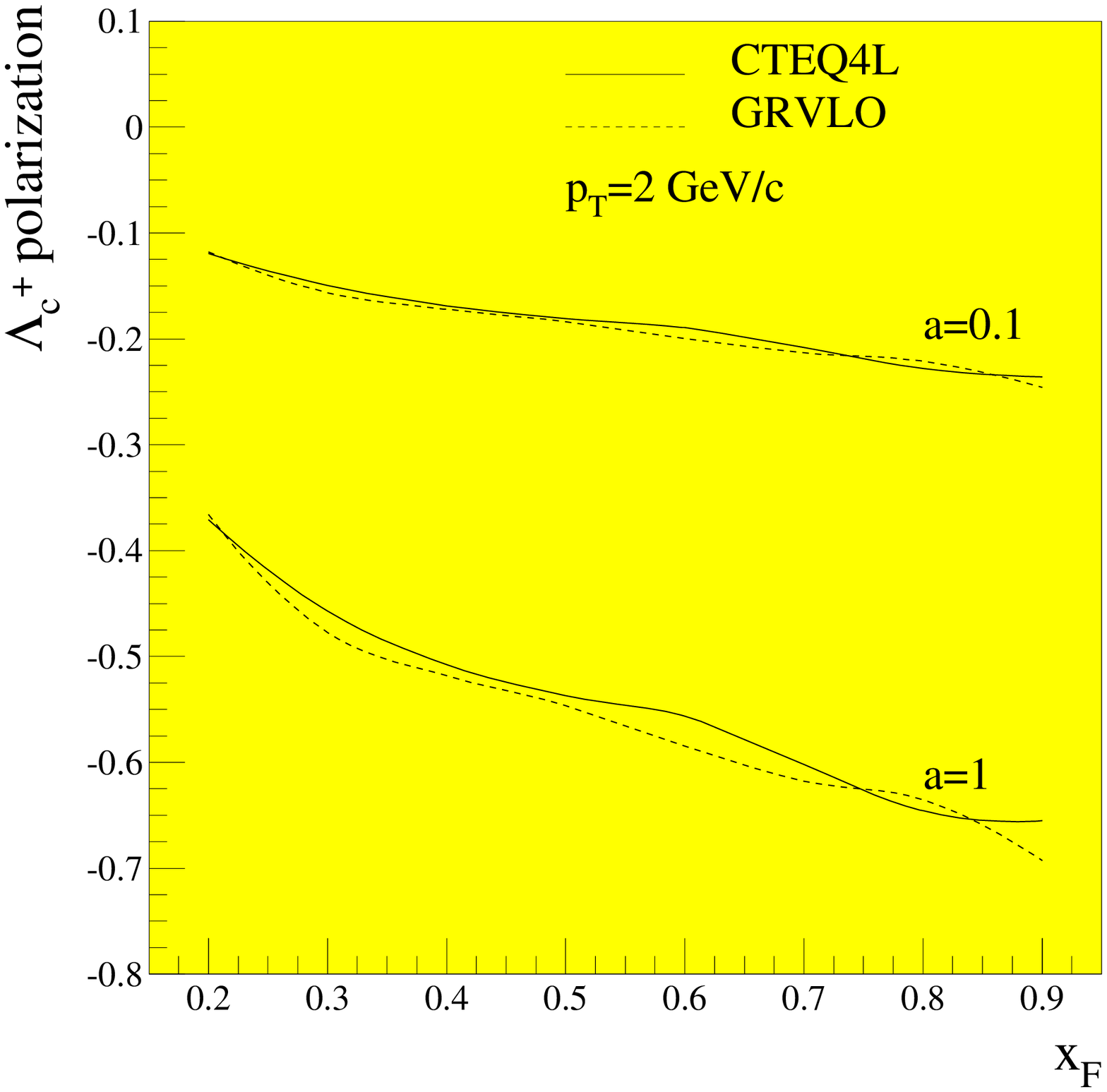,width=12cm,height=8cm}\\
\vspace*{4mm}
\hspace*{1.5cm} \parbox{12cm}{
  Fig. 4. $x_F$ dependence of the $\Lambda_c^+$ polarization in the process
$p \,+\, {\rm Pomeron} \to \Lambda_c^+ \,+\, X$ at $p_T$\,=\,2\,GeV/c.
The solid (dashed) line corresponds to CTEQ4L(GRVLO) parton distribution 
functions.} 
\end{center}

\newpage
\begin{thebibliography}{99}
\bibitem {Heller} (a) Heller K., in {\it Proc. of 12th Int. High Energy Spin 
Physics Symp., Amsterdam, 1996}, edited by deJager C.W. {\it et al.}, p. 23;
(b) strange baryon polarizations were also reviewed by  Bravar A.
at {\it 13th Int. High Energy Spin Physics Symp.}, Protvino, Sept, 1998.
\bibitem {Lubimov} Aleev A.N. {\it et al., Yad.Fiz.}, 1986, vol.43,p.619.
\bibitem {close} Close F.E. and Roberts R.G., {\it Phys. Lett.} 1994, 
vol. B336, p.257;
Goloskokov S.V., {\it Phys.Lett.} 1993, vol. B315, p.459; 
Goloskokov S.V., in ref. [1a], p. 334.
\bibitem {YuA1} Arestov Yu.I. {\it et al., Sov.\,J.\,Nucl.\,Phys.} 1985,
vol. 42, p.483;
 {\it Yad.\,Fiz.} 1985,vol. 42, p.761; \\
also in Arestov Yu.I. and Nurushev S.B., {\it Exchange degeneration and 
polarization in inclusive fragmentation processes $p \to \Lambda K$ and 
$p \to \Lambda_c \bar D$}, IHEP 83-125, 1983 (unpublished).
\bibitem {Simao} dos Anjos J., Herrera G., Magnin J. and Sima\~o F.R.A.,
{\it Phys.\,Rev.} 1997, vol. D56, p. 394; {\it e-print} hep-ph/9702257.
\bibitem {PDG} Particle Data Group, {\it Phys. Rev.} 1996, vol. D54, Part 1, 
Review of Particle Physics.
\bibitem {YuA2} Arestov Yu., {\it Twist-3 induced asymmetries in hard 
production at 70 GeV/c}, in ref. [1a], p. 187.
\bibitem {Szwed} Szwed J., {\it Phys.\,Lett.} 1981, vol. 105B, p. 403.
%
\bibitem {YuA3} Amaglobeli M.S. et al., {\it Sov.\,J.\,Nucl.\,Phys.}
1989, vol. 50, p. 432; {\it Yad.\,Fiz.} 1989, vol. 50, p. 695.
%
\bibitem {pdf}   CTEQ4: Lai  H.L. {\it et al., Phys. Rev.} 1997, vol. D55,
p. 1280; {\it e-print} hep-ph/9606399; \\
 GRVLO: Gl\=uck M., Reya E. and Vogt A., {\it Z.\,Phys.} 1995, vol. C67,
p. 433.

\end{thebibliography}
\end{document}